\documentclass[reprint,amsmath,amssymb,aps,pra,notitlepage,twocolumn]{revtex4-1}
\usepackage{graphicx}
\usepackage{dcolumn}
\usepackage{bm}

\begin{document}


\title{The influence of the interaction between
quasiparticles on parametric resonance in
Bose-Einstein condensates quasicondensates}



\author{Maciej Pylak}
\affiliation{National Centre for Nuclear Research, ul. Ho\.za 69, 00-681 Warsaw, Poland}

\author{Pawe{\l} Zin}
\email{Pawel.Zin@ncbj.gov.pl}
\affiliation{National Centre for Nuclear Research, ul. Ho\.za 69, 00-681 Warsaw, Poland}


\begin{abstract}
We perform a simulation of the experiment \cite{Paryz1} where the temporal modification
of the effective one dimensional interaction constant was used to 
create pairs of atoms with opposite velocities.
The simulations clearly demonstrate  huge impact of interaction between quasiparticles
due to finite temperature on the pair production process, explaining 
relatively small atom pair production and the absence of the number squeezing in the experiment.
\end{abstract}

\maketitle

\newcommand{\x}{{\bf r}}
\newcommand{\K}{{\bf k}}
\newcommand{\dk}{  \Delta {\bf k}}
\newcommand{\DK}{\Delta {\bf K}}
\newcommand{\KK}{{\bf K}}
\newcommand{\X}{{\bf R}}

\newcommand{\B}[1]{\mathbf{#1}} 
\newcommand{\f}[1]{\textrm{#1}} 

\newcommand{\half}{{\frac{1}{2}}}

\newcommand{\vv}{{\bf v}}
\newcommand{\p}{{\bf p}}

\section{Introduction}\label{Int}

The generation of non-classical states in atomic ensembles is a rapidly developing 
 direction in the trapped ion and cold neutral atomic physics \cite{przeglad}.
Such states can be used to increase the sensitivity of precision measurements
beyond the standard classical limit \cite{przeglad2}.
A hundred times decrease of measurement noise beyond the classical limit 
was recently reported  in cold thermal atoms \cite{Kasevich}.
One of the possible states that are particle entangled, and can be used to 
increase the sensitivity of precision measurements is a so-called twin-Fock state $|n,n\rangle$ \cite{limit1,limit2}.
Such a state can be created in experiments generating atomic pairs with well defined momenta in quasi one-dimensional systems.
This was done by modulation of the effective one-dimensional atomic interaction parameter
\cite{Paryz1}, or a modulation instability present in a one-dimensional lattice \cite{Paryz2},
or else by the decay of an excited state \cite{Wieden}.
The theoretical analysis for these situations  was performed using the Bogoliubov approximation \cite{Wasak,instability,Casimir}.
In this case, the Hamiltonian is quadratic in field operators. 
It has the term responsible for the creation of atomic pairs but neglects the terms of higher order in field operators, which describe the interaction between quasiparticles.
As the process of pair creation starts, the atoms, according to the Bogoliubov description,
are created in pairs with well defined momenta.
Therefore, one expects violation of Cauchy-Schwartz inequality,
which is the clear signature of entanglement  \cite{jan}. 
In two of the mentioned experiments such violation  was observed \cite{Paryz2,Wieden}.
However, it was not seen in the experiment described in \cite{Paryz1}.
This suggests that the interaction between quasiparticles, neglected in the Bogoliubov approximation,
can influence the pair production.
We performed such analysis for three dimensional homogeneous system in the case of pair creation
caused by temporal modulation of the interaction parameter \cite{nasza}.
There was found that indeed the interaction between quasiparticles may drastically change the pair creation process.
The parametric process is described by a single parameter $\delta$  responsible for the strength of the amplification, while the interaction between quasiparticles is described by a quasiparticle decay constant $\gamma$.
When $\delta > \gamma$ the pair production process is roughly given by $\exp \left( 2 (\delta - \gamma) t \right)$
leading always to huge number of pairs produced if $t$ is large enough.
On the other hand if $\gamma > \delta$ the number of pairs produced tends to a constant for $t \rightarrow \infty$.
Additionally we have found that depending on the parameters of the system the Cauchy-Schwartz inequality may be, or may be not violated.

In the present paper we perform numerical simulation of the experiment \cite{Paryz1}
using classical field method \cite{clas}. 
The obtained results agree with the experimental measurements showing relatively small
pair production and the lack of Cauchy-Schwartz equality violation. 
To get deeper understanding of the obtained results we perform numerical simulation of the
homogeneous analogue of the experimental system. There we additionally perform Bogoliubov's 
method based analysis.  
For the temperatures much smaller than in the experiment, the classical field method results
agree with the Bogoliubov method predictions showing huge number of pairs produced.
But when the temperature tends to the experimental value the production process
practically stops with relatively small number of atom pairs produced which is in agreement
with the experiment.
To check if the condition derived in \cite{nasza} applies to the one-dimensional case, 
we  numerically compute  $\gamma$ as well as the number of atom pairs produced as
a function of temperature. We find that the number of pairs produced tends to a constant
if  $\gamma$ is approximately equal to $\delta$ which validates the suggested condition.

The plan of the paper is as follows.
In Section \ref{sec1} we describe the system of interest.
There we introduce classical field method as an approximate method of description.
In Section \ref{sec2} using Bogoliubov approximation to the classical field method
we construct initial thermal state of the system. 
In Section \ref{sec3} we siimulate the pair creation process
for both homogeneous and inhomogeneous systems.
We conclude in Section \ref{sec4}.

\section{Theoretical model}\label{sec1}

We consider the system described in \cite{Paryz1}.
There we have $N=10^5$ helium atoms of temperature $T=200$nK put in harmonic trapping potential
$V(\x) = \frac{1}{2}m [ \omega_r^2(y^2+z^2) + \omega_x^2 x^2]$ with
$\omega_r = 1500 \times 2 \pi$~Hz and $\omega_x = 7 \times 2\pi$~Hz. The atoms interact via
contact potential characterized by parameter $g_{3d} = \frac{4 \pi \hbar^2 a}{m}$
where $a=7.51$~nm is the  metastable helium scattering length.
In the experiment the laser intensity  oscillates in time what causes 
oscillations of trapping frequencies. 
When oscillation ends the trapping potential is turned off,
the atoms freely expand and finally fall on the detector. The measurement of a time of arrival and a position of the atom detected allows for a reconstruction of the atomic velocity correlation functions.

\begin{figure}[htb]
	\centering
		\includegraphics[width=0.48\textwidth]{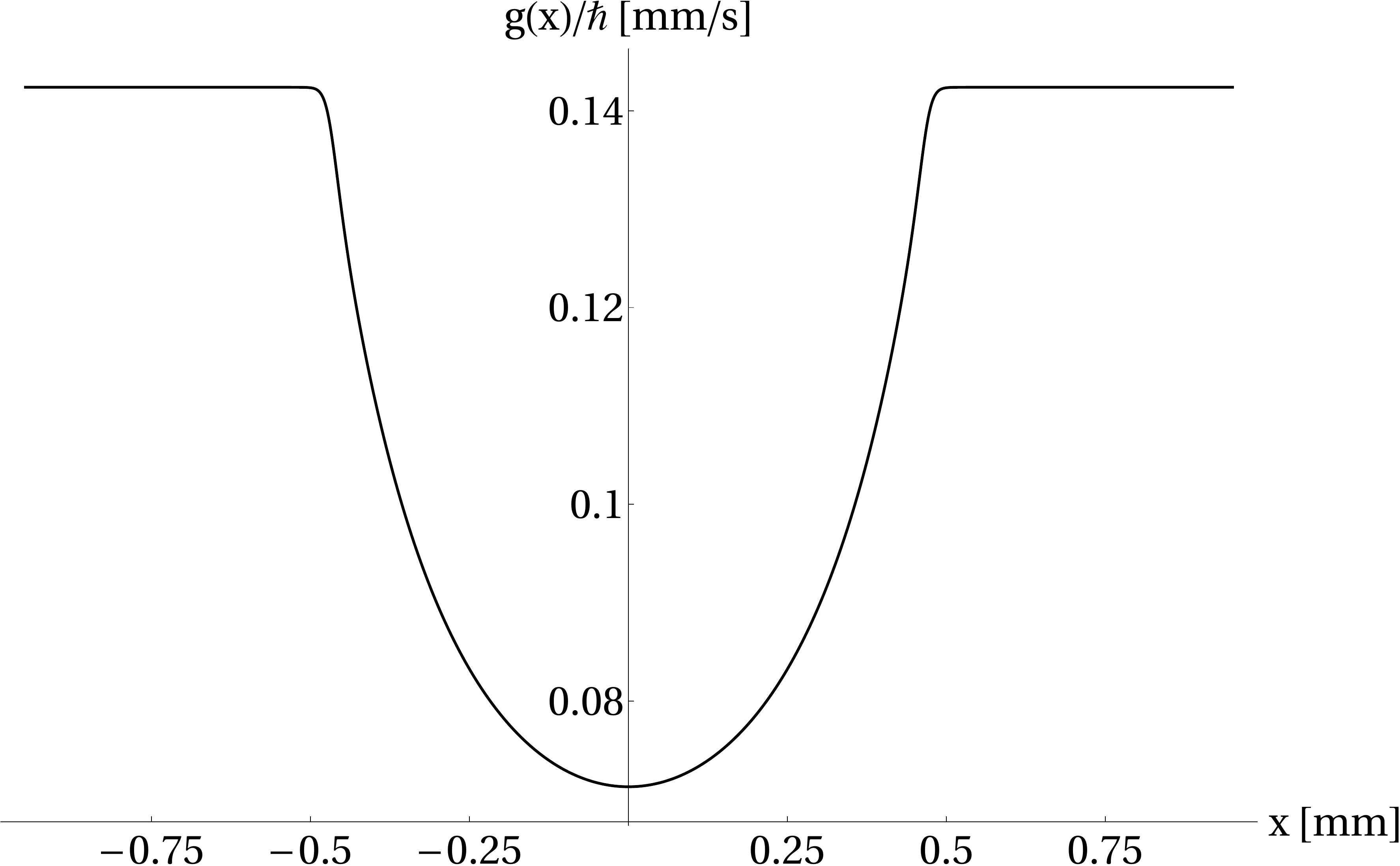}
		\caption{Effective 1D-interaction parameter.}
		\label{fig:gx}
\end{figure}

Let us now introduce the theoretical model of the above experimental scenario.
As $\omega_x \ll \omega_r$ the system is highly elongated. 
Therefore we describe the system in an approximate manner by the one-dimensional gas of atoms interacting via effective parameter 
\begin{equation}\label{g1d}
g(x) = g_{3d}\frac{\int{\mbox{d}y \mbox{d}z \, n_{3d}^2(\x)}}{\left(\int{\mbox{d}y \mbox{d}z \, n_{3d}(\x)}\right)^2}
\end{equation}
where $n_{3d}(\x)$ is given by the solution of stationary Gross-Pitaevskii equation
\begin{equation}\label{GP3d}
\left( - \frac{\hbar^2}{2m}\Delta + V(\x) + g_{3d} n_{3d}(\x)  - \mu_{3d} \right) \sqrt{n_{3d}(\x)} = 0
\end{equation}
with the normalization condition $\int{\mbox{d} \x \, n_{3d}(\x)} = N$.
The function $g(x)$ obtained using above equations is plotted in Fig.~(\ref{fig:gx}).
One can clearly see the position dependence of the interaction parameter.
As we shall see later, oscillation of the trapping frequencies causes the effective  interaction constant to be time dependent $g(x,t)$.
 We further approximate the description of the system by relying on classical field method.
This approximation is allowable if the system is weakly interacting i.e. if $\gamma \ll 1$
where
\begin{equation} \label{gamma}
 \gamma = \frac{m g}{\hbar^2 n}
\end{equation}
and $n$ being 
the one-dimensional particle density. For our system we can approximate 
$n(x) \simeq \int \mbox{d}y \mbox{d}z \, n_{3d}(\x) $ which gives $n(0) \simeq 1.8 \times 10^8$atoms/m
and   $\gamma \simeq  \frac{m g(0)}{\hbar^2 n(0)} \simeq 2.5 \times 10^{-5}$ which justifies the use of classical field approximation. 
In this method we substitute  creation and annihilation operators of highly populated modes
with c-numbers $\hat a_\nu \rightarrow \alpha_\nu$.
To perform this approximation we approximate the continuous space by finite lattice.
The Hamiltonian of the system takes the form
\begin{eqnarray} \nonumber
\hat H &=& \sum_x \Delta x  \, \hat \Psi^\dagger(x) \left( - \frac{\hbar^2}{2m}\triangle_d  + V(x) \right)  \hat \Psi(x)
\\
&+& \sum_x  \Delta x \, \frac{g(x,t)}{2} \hat \Psi^\dagger(x) \hat \Psi^\dagger(x) \hat \Psi(x) \hat \Psi(x) 
\end{eqnarray}
where the sum is over discrete points $x = \frac{j}{M} L$, $j=1,...,M$ with $L$ being the length of the system and
$\Delta x = \frac{L}{M}$. Additionally $\triangle_d$ is a finite matrix approximation of the one-dimensional laplacian that we shall specify later on. We consider harmonic external potential $V(x) = \frac{1}{2}m \omega^2 x^2$ and, as mentioned above,  time and position dependent interaction parameter  $g(x,t)$.
The commutation relation takes the form $[\Psi(x),\Psi^\dagger(y)] = \frac{\delta_{x,y}}{\Delta x}$ and lead to
Heisenberg equation of motion:
\begin{eqnarray} \nonumber
i \hbar \partial_t \hat \Psi(x,t) &=& \left( - \frac{\hbar^2}{2m} \triangle_d  + V(x) \right) \hat \Psi(x,t)
\\ \label{GPH}
&+& g(x,t) \hat \Psi^\dagger(x,t) \hat \Psi(x,t) \hat \Psi(x,t).
\end{eqnarray}
In the classical field approximation the field operator $\hat \Psi(x,t) $
turns into classical field $\psi(x,t)$. 
Then (\ref{GPH}) becomes well known Gross-Pitaevskii (GP) equation
\begin{eqnarray} \nonumber
i \hbar \partial_t \psi(x,t) &=& \left( - \frac{\hbar^2}{2m} \triangle_d  + V(x) \right) \psi(x,t)
\\ \label{GP}
&+& g(x,t)  |\psi(x,t)|^2  \psi(x,t).
\end{eqnarray}
The quantum state is now replaced by probability distribution of classical field values $P[\psi(x,t)]$
and the quantum averages are now substituted by averages over this probability distribution.
Initially the system is in thermal equilibrium. 
Since it is isolated one should use the microcanonical ensemble to describe
the state of the system,  thus  $P[\psi(x,0)]$ is the microcanonical ensemble
probability distribution. Having $P[\psi(x,0)]$ we can calculate  mean value of any observables in
$t=0$. For $t>0$ we proceed in the following way.
We  draw a single random $\psi(x,0)$  from $P[\psi(x,0)]$ which gives us  
single realization of $\psi(x,0)$. 
Then each realization is evolved using (\ref{GP}) to the final time.
The observables are then calculated as the averages over the realizations.
One of the parameters of control in the microcanonical ensemble is the energy of the system.
However, in the experiment the temperature is the measured parameter.
Thus we need to generate single realizations $\psi(x,0)$ for a given temperature.
This seems to be a complicated task and we have chosen another, approximate way 
of obtaining initial state with desired temperature which we described in the following Section.

\section{ Temperature diagnostics and initial state preparation}\label{sec2}

For temperatures low enough the weakly interacting one-dimensional gas enters the quasicondensate regime, where it is described as a system of  weakly interacting Bogoliubov quasiparticles \cite{gora}.
There the classical field (in the quantum description the field operator) is
represented as
\begin{equation}\label{psin}
\psi(x,0) = \sqrt{n(x) + \delta n(x)} e^{i \phi(x)}.
\end{equation}
where 
\begin{eqnarray} \label{dn}
 \delta  n(x) &=& \sqrt{n(x)} \sum_\nu f_\nu^-(x)  \alpha_{\nu} + \mbox{c.c.}
\\ \label{phi}
\phi(x) &=& \frac{1}{\sqrt{4 n(x)}} \sum_\nu  - if_\nu^+(x)  \alpha_{\nu} + \mbox{c.c.}.
\end{eqnarray}
In the above $\alpha_\nu$,  $\hbar \omega_\nu$ and $ f_{\nu}^{\pm}$ are mode amplitudes, energies and mode functions respectively, 
obtained via solution of Bogoliubov--de Gennes equations
\begin{eqnarray}\nonumber
&& \left( H_0(x) - \hbar \omega_\nu \right) u_\nu(x) - g(x)n(x) v_\nu(x) = 0
 \\ \label{bog2}
\\
&& \left( H_0(x) + \hbar \omega_\nu \right) v_\nu(x) - g(x)n(x) u_\nu(x) = 0
\end{eqnarray}
where  $f_\nu^\pm = u_\nu \pm v_\nu$  are normalized by condition 
\begin{equation}\label{co}
\sum_x \Delta x \, (f^+_\nu(x))^* f^-_{\nu'}(x) = \delta_{\nu,\nu'}. 
\end{equation}
Here $n(x)$ is given by the solution of the stationary GP equation
\begin{equation}\label{sGP}
H_0(x) \sqrt{n(x)} = 0 
\end{equation}
with the normalization condition $\sum_x \Delta x \, n(x) = N$, where
\begin{eqnarray*}
H_0 = -\frac{\hbar^2}{2m} \triangle_d + V(x) + 2 g(x) n(x) - \mu.
\end{eqnarray*}
Neglecting the weak interaction between quasiparticles the 
Hamiltonian of the system is $H \simeq \sum_\nu \hbar \omega_\nu |\alpha_\nu|^2$.
The above description motivates us to approximate the probability distribution $P[ \psi(x),0] $ 
by the one corresponding to the canonical ensemble of noninteracting quasiparticles, that is
$P[\psi(x),0] =  \prod_\nu  P(\alpha_\nu)$ where
\begin{equation} \label{Pcl}
P(\alpha_\nu) = \frac{\hbar \omega_\nu}{\pi k_BT_{in}} \exp \left(  - \frac{\hbar \omega_\nu}{k_bT_{in}} |\alpha_\nu|^2 \right)
\end{equation}
where $T_{in}$ denotes initial temperature.

We now specify the finite matrix approximation to one-dimensional laplacian $\triangle_d$
taking the one given by the discrete Fourier transform 
\begin{eqnarray*}
&& - \triangle_d(x,x_k) = \frac{1}{M} \sum_k  k^2 e^{ik(x-x_k)}  
\\
&& =   \left(\frac{2\pi}{L} \right)^2 
\left(\frac{(-1)^x}{2 \sin^2 \left( \frac{\pi}{M} x \right)} (1-\delta_x)  + \delta_x C \right)
\end{eqnarray*}
where $k = 2 \pi \frac{m}{M}$, $x = \frac{x-x_k}{\Delta x}$, $  -\frac{M}{2} +1 \geq m \leq \frac{M}{2}$ 
and $C = \frac{1}{M}\sum_{m= -M/2 +1}^{M/2} m^2 = \frac{1}{12}(M^2+2) $. By doing so we implicitly assume periodic boundary
condition.
Having specified $\triangle_d$ and knowing $g(x)$ we solve the one-dimensional GP equation (\ref{sGP})
obtaining  Thomas-Fermi radius equal to $R=0.46$~mm, 
chemical potential $\frac{\mu}{\hbar} = 2.05 \times 2 \pi$~kHz and  the maximal density of the system
equal to $n \simeq 1.8 \times 10^8$~atoms/m. 
Before further analysis of the nonuniform system let us first analyse its homogeneous analogue.

\subsection{Homogeneous system}

The parameters of the homogeneous system were chosen in the following way.
We take the interaction parameter $g=g(0)$ and the mean density equal to the maximal density of the trapped system so that the chemical potentials of both uniform and nonuniform systems are the same.
The box size is chosen to be equal to $L=2R = 0.92$~mm with $M=1024$ points. 
In the homogeneous case the solution of the Bogoliubov--de Gennes equations takes the analytical form \cite{gora}
\begin{equation}\label{fhom}
 f_k^\pm(x) = \frac{1}{\sqrt{L}} \left( \frac{\hbar \omega_k}{E_k} \right)^{\pm 1/2} e^{ikx} = f_k^\pm e^{ikx}
\end{equation}
where
\begin{equation} \label{boghom}
 \hbar \omega_k = \sqrt{E_k(E_k + 2 ng)}
\end{equation}
and $E_k = \frac{\hbar^2 k^2}{2m}$.
We additionally calculate the density fluctuations equal to
\begin{eqnarray*}
&&\frac{\langle \delta n^2 \rangle}{n^2} = \frac{2}{n} \sum_k (f_k^-)^2 \langle |\alpha_k|^2 \rangle 
\\
&&
=  \frac{k_B T}{n \pi} \int \mbox{d} k \, \frac{1}{E_k + 2 ng} =  \frac{k_B T}{ng} \sqrt{ \gamma } 
\end{eqnarray*}
where $\gamma$ is given by Eq.~(\ref{gamma}). To obtain the above we used Eqs.~(\ref{dn}), (\ref{fhom}),
(\ref{boghom}) and (\ref{Pcl}).
For the parameters given above and $T_{in}=200$~nK  we obtain $\sqrt{\langle \delta n^2 \rangle }/n \simeq 0.01$.
The fact that it is much smaller than unity justifies the use of Bogoliubov method.

The single realization of the initial state $\psi(x,0)$ is constructed using Eqs.~(\ref{dn}), (\ref{phi}) and  (\ref{psin})
upon drawing $\alpha_\nu$ randomly from the distribution (\ref{Pcl}) with chosen $T_{in}$.
Still this is not the end of the construction.
However, as we have written above, the distribution  (\ref{Pcl}) neglects the interaction between quasiparticles.
To get the initial state we proceed in the following way.
We evolve just constructed state using GP equation (\ref{GP}) with $g(x,t)= g(x)$
and $V(x)=0$ for sufficiently long time.
After some time due to the interaction between quasiparticles present in the GP equation 
the system thermalizes and reaches its equilibrium state.
In fact we do not know the temperature of this state.
We find its approximate value $T$ using a procedure described below based on Bogoliubov method.
If $T \simeq T_{in}$ then the influence of interaction between quasiparticles on the temperature 
is negligible and it is justified to state that the temperature of the prepared state
equals to $T_{in}$.

The approximate procedure to extract the temperature uses the 
decomposition
\begin{eqnarray} \label{dnt}
 \delta  n(x,t) &=& \sqrt{n(x)} \sum_k f_k^-(j)(x)  \alpha_k(t) + \mbox{c.c.}
\\ \label{phit}
\phi(x,t) &=& \frac{1}{\sqrt{4 n(x)}} \sum_k  - if_k^+(x)  \alpha_k(t) + \mbox{c.c.}.
\end{eqnarray}
where $|\alpha_k(t)|$ are time dependent due to interaction  between quasiparticles omitted in
the Bogoliubov approximation. Using orthogonality condition (\ref{co}) together with the above 
equations we obtain
\begin{equation}\label{alphat}
\alpha_k(t) = \sum_x \Delta x \, \frac{1}{\sqrt{L}} e^{-ikx} 
\left( f_k^+\frac{\delta n(x,t)}{2\sqrt{n}} + i f_k^- \sqrt{n} \phi(x,t) \right)
\end{equation}
where we used the fact that $f_k^\pm$ are real. 
In every single realization having $\psi(x,t)$  we find $\delta n(x,t)$
and $\phi(x,t)$ and upon inserting it into (\ref{alphat}) we obtain all $\alpha_k(t)$.
This enables us to find the average over many realizations
 $\langle |\alpha_k|^2 \rangle = \frac{1}{N_r} \sum_{i=1}^{N_r} |\alpha_{k,r}|^2 $
where $r$ denotes the number of realization.
We assume the distribution of $\alpha_k$ to be given by the formula 
(\ref{Pcl}) with $T_{in}$ changed into k-dependent final temperature $T_k$.
Having that we derive equipartition formula $\langle |\alpha_k|^2 \rangle  = \int \mbox{d}^2 \alpha_k \, |\alpha_k|^2 P(\alpha_k)  = \frac{k_B T_k}{\hbar \omega_k}$ which connects numerically calculated averages $\langle |\alpha_k|^2 \rangle$
with the final temperature $T_k$.
\begin{figure}[htb]
	\centering
		\includegraphics[width=0.48\textwidth]{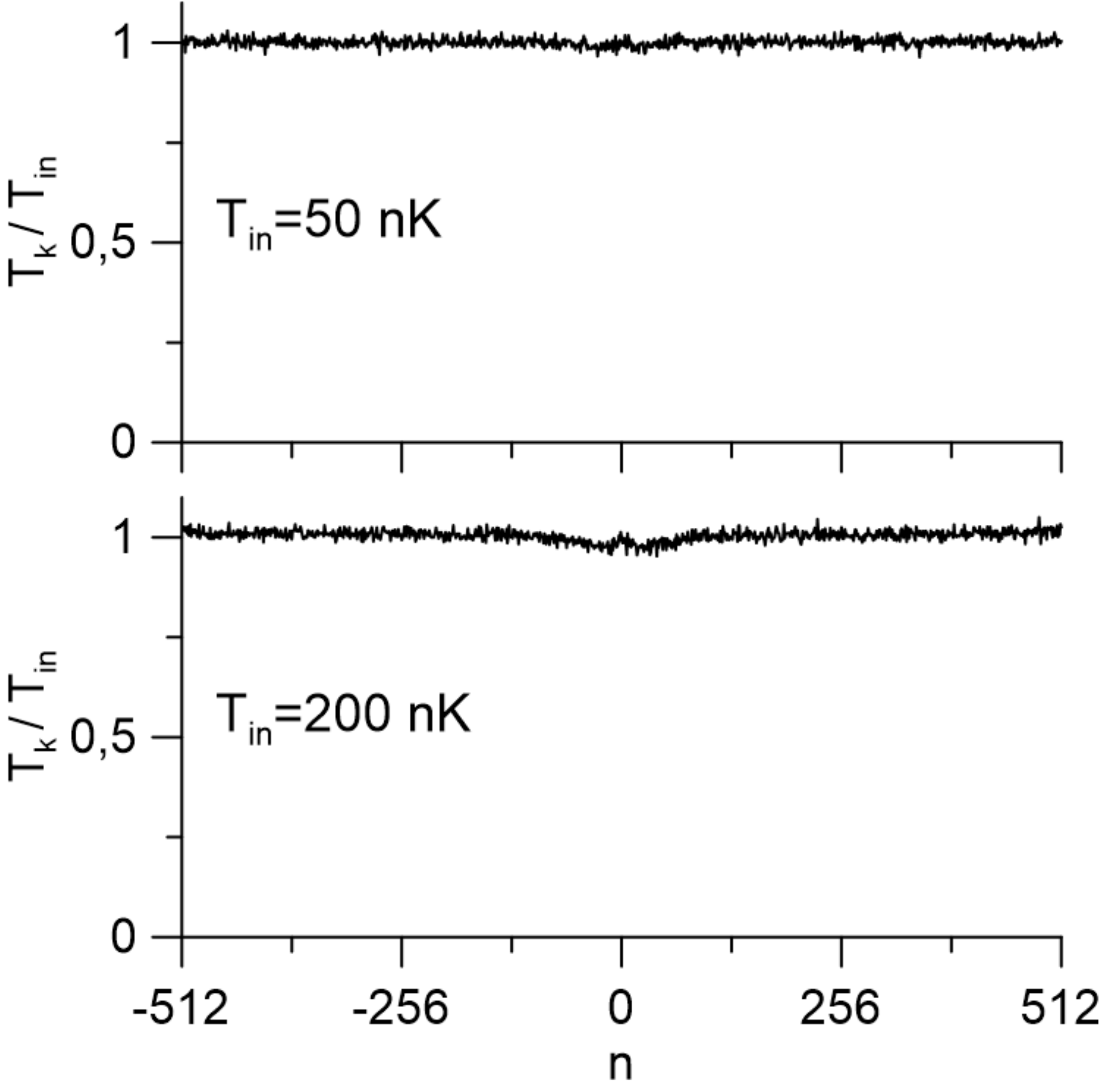}
		\caption{The ration $T_k/T_{in}$ as the function of $n$ where $k=2\pi n/L$ for 
		$T_{in} = 50,200$ nK.}
		\label{Tk}
\end{figure}
In Fig.~(\ref{Tk}) we plot the ratio $T_k/T_{in}$ as the function of $n$ where $k=\frac{2\pi n}{L}$ for $T_{in}=50,200$ nK. 
We clearly see that  $T_k$ fluctuates around the value being very close to the initial temperature $T_{in}$.
That shows that the temperature of the thermalized state is equal to $T_{in}$.

\subsection{Inhomogeneous system}

As in the homogeneous case we would like to use here Bogoliubov method to prepare
the initial thermal state with given temperature $T_{in}$.
However, the Bogoliubov method described above is a correct approximation
in the bulk region of the system where $\delta n/n \ll 1$.
At the edges of the system where $n$ is negligible and thermal cloud dominates,
the condition $\delta n/n \ll 1$ is no longer satisfied and the Bogoliubov method cannot be used. 
Still we need to construct classical field $\psi(x,0)$ in that region.
Below we describe the approximate method to do that.

We divide our system into three parts: the bulk region
$|x| < x_0$ ( $x_0 < R$ ) where the condition  $\delta n/n \ll 1$ 
is satisfied and where we use Bogoliubov description.
The region outside the quasicondensate $|x|> R$ where $n$ is practically equal to zero.
As we checked numerically in that region $v_\nu(x) \simeq 0$ and
 the Bogoliubov equations (\ref{bog2}) take an approximate form 
\begin{eqnarray*}
\left( - \frac{\hbar^2}{2m} \triangle_d + V(x)  \right) u_\nu(x) = (\mu + \epsilon_\nu)  u_\nu(x).
\end{eqnarray*}
The above equation is not a surprise since in that region we in fact neglect the nonlinear term
$g|\psi|^2 \psi$ in the GP equation (\ref{GP}) ending up with noninteracting gas.
Then the classical field simply equals to 
\begin{equation} \label{psi000}
 \psi_\mp (x,0) =  e^{i \phi_{\mp}} \sum_\nu u_\nu(x) \alpha_\nu 
\end{equation}
where $\psi_\mp$ denotes the field in the $x < - R$ and $x > R$ regions.
Above we have also introduced phases $\phi_{\mp}$ for the reason which
shall become clear later.
The last region is  $ x_0 < |x|< R$ where $n(x) \neq 0$ but the 
 condition $\delta n /n \ll 1$ is not satisfied.
In that region we take
\begin{equation}\label{psimp2n}
\psi_\mp(x) =  A_\mp(x) e^{i\phi_\mp}  \left( \sqrt{n(x)} + \delta \psi(x) \right)
\end{equation}
where
\begin{equation}\label{deltapsi}
\delta \psi = \sum_{\nu } \left(u_\nu \alpha_\nu - v_\nu \alpha_\nu^*\right).
\end{equation}
We introduce  $A_\mp(x) e^{i\phi_\mp}$ factor where we take $A_\mp$ as a real function 
to match the continuity conditions at $|x| = x_0$ which take the form
\begin{eqnarray*}
&& A_\mp(\mp x_0) e^{i\phi_\mp} \left( \sqrt{n(\mp x_0)} + \delta \psi(\mp x_0) \right)
\\
&& = \sqrt{n(\mp x_0)  + \delta n(\mp x_0)} e^{i\phi(\mp x_0)}.
\end{eqnarray*}
Using Eq.~(\ref{deltapsi}) and the fact that in the region $|x|> R $, $v_\nu(x) \simeq 0$
we rewrite Eq.~(\ref{psi000}) as
\begin{equation} \label{psimp}
 \psi_\mp (x,0) =  e^{i \phi_{\mp}} \delta \psi(x,0).
\end{equation}
Then the continuity condition at $x = \mp R$ where $n(\pm R) =0$ 
implies $A_\mp(\mp R) = 1$.
To fully define the classical field we need to specify the function $A(x)$
between $|x|=x_0$ and $|x|=R$. We make the simplest choice of linear function
\begin{eqnarray*}
&& A_\mp(x)  =  \theta(|x| - x_0) 
\\
&&
+ \theta(x_0- |x|) 
\left( A_\mp(x) \frac{R-|x|}{R-x_0} +  \frac{|x|-x_0}{R-x_0} \right).
\end{eqnarray*}
The single realization of the classical field $\psi$ consists of drawing $\alpha_\nu$ from the distribution (\ref{Pcl})
and inserting it into Eqs.~(\ref{dn}), (\ref{phi}) and (\ref{psin}) to get the bulk region part
and in  Eqs.~(\ref{deltapsi}), (\ref{deltapsi}) and (\ref{psimp}) to get the tails of $\psi$.

\begin{figure}[htb]
	\centering
		\includegraphics[width=0.45\textwidth]{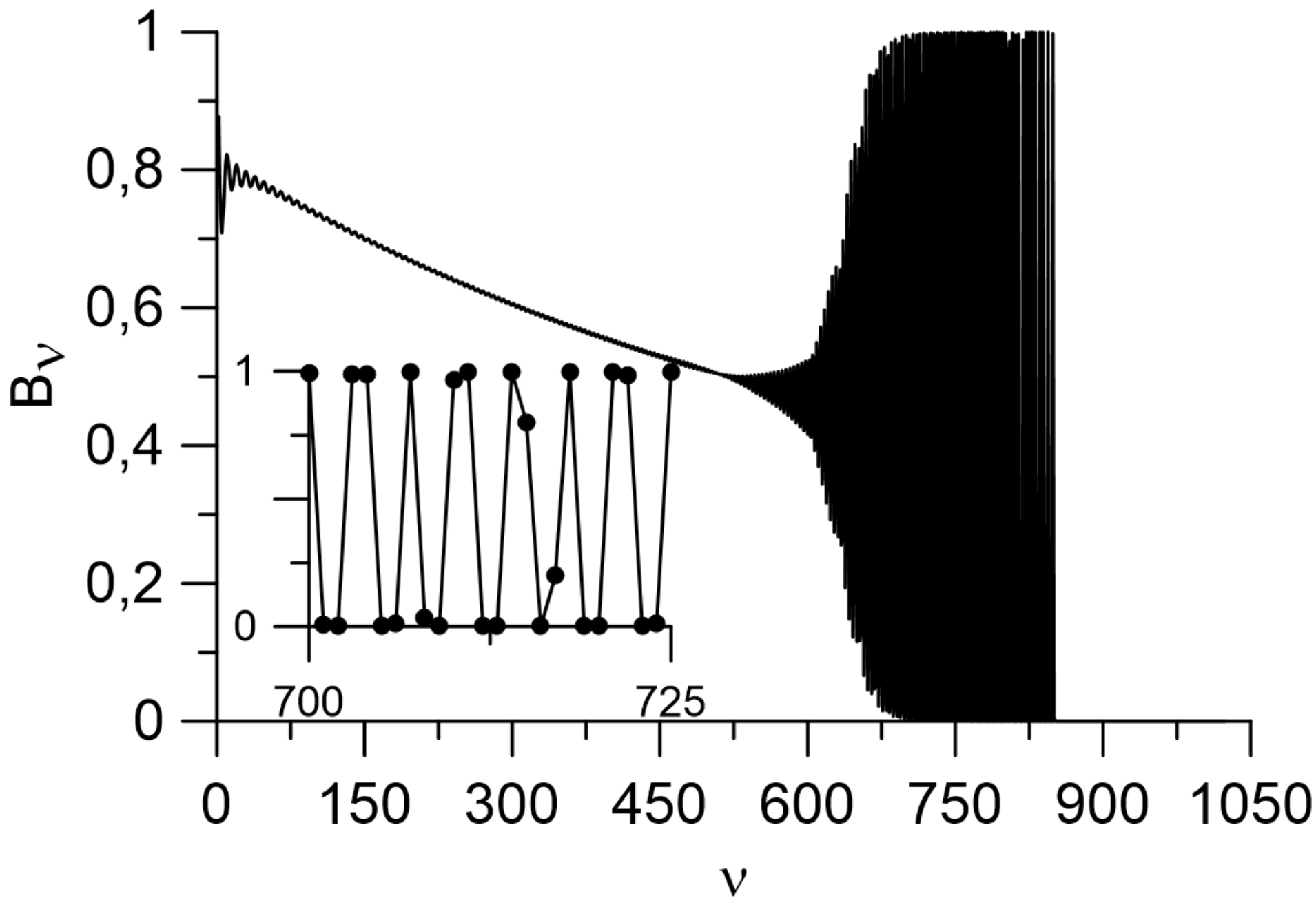}
		\caption{The function $B_\nu$. In the inset we plot the magnification
			in the smaller range to show the rapid oscillation.}
		\label{fig:bv}
\end{figure}

\begin{figure}[htb]
	\centering
		\includegraphics[width=0.48\textwidth]{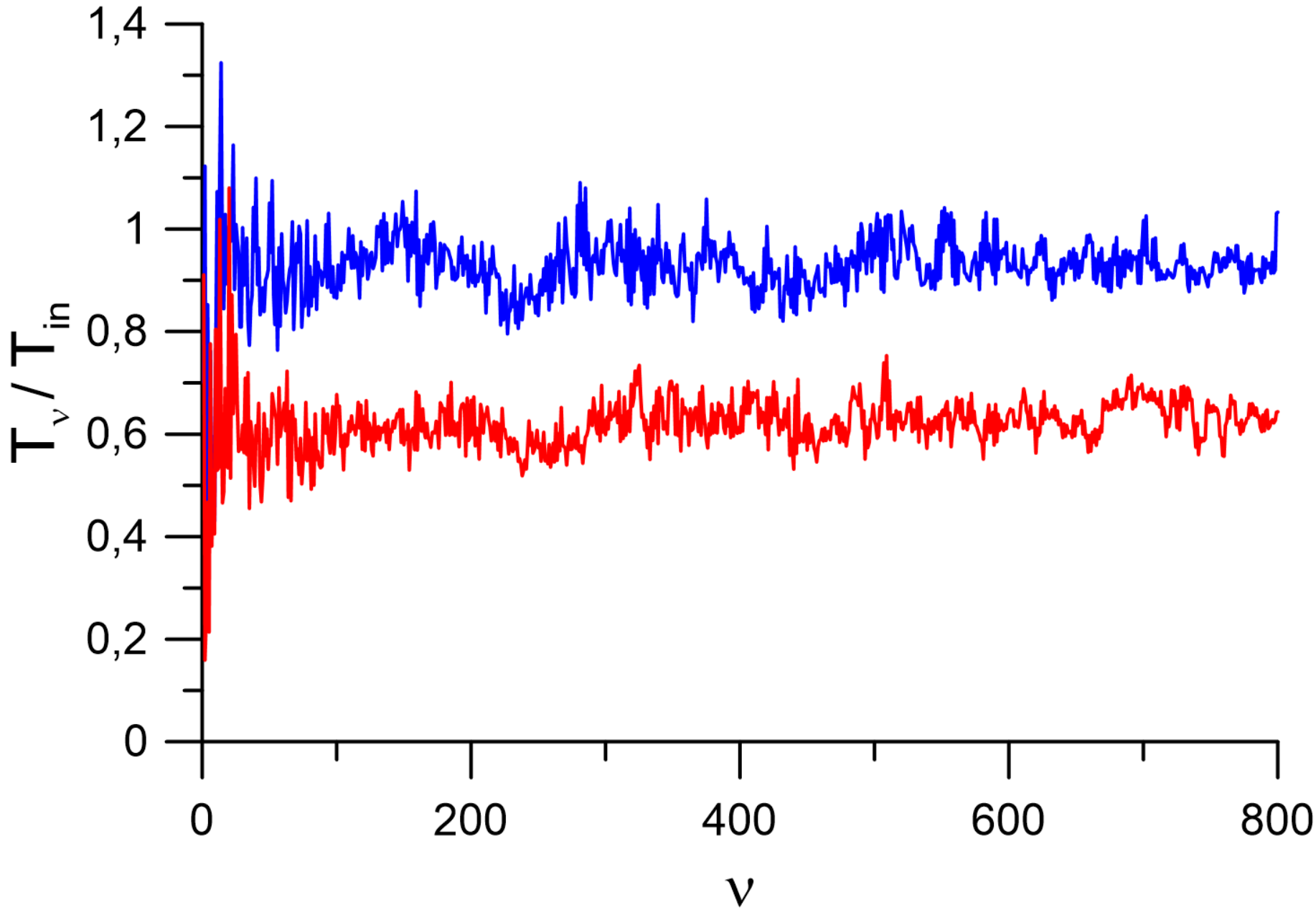}\\
		\includegraphics[width=0.48\textwidth]{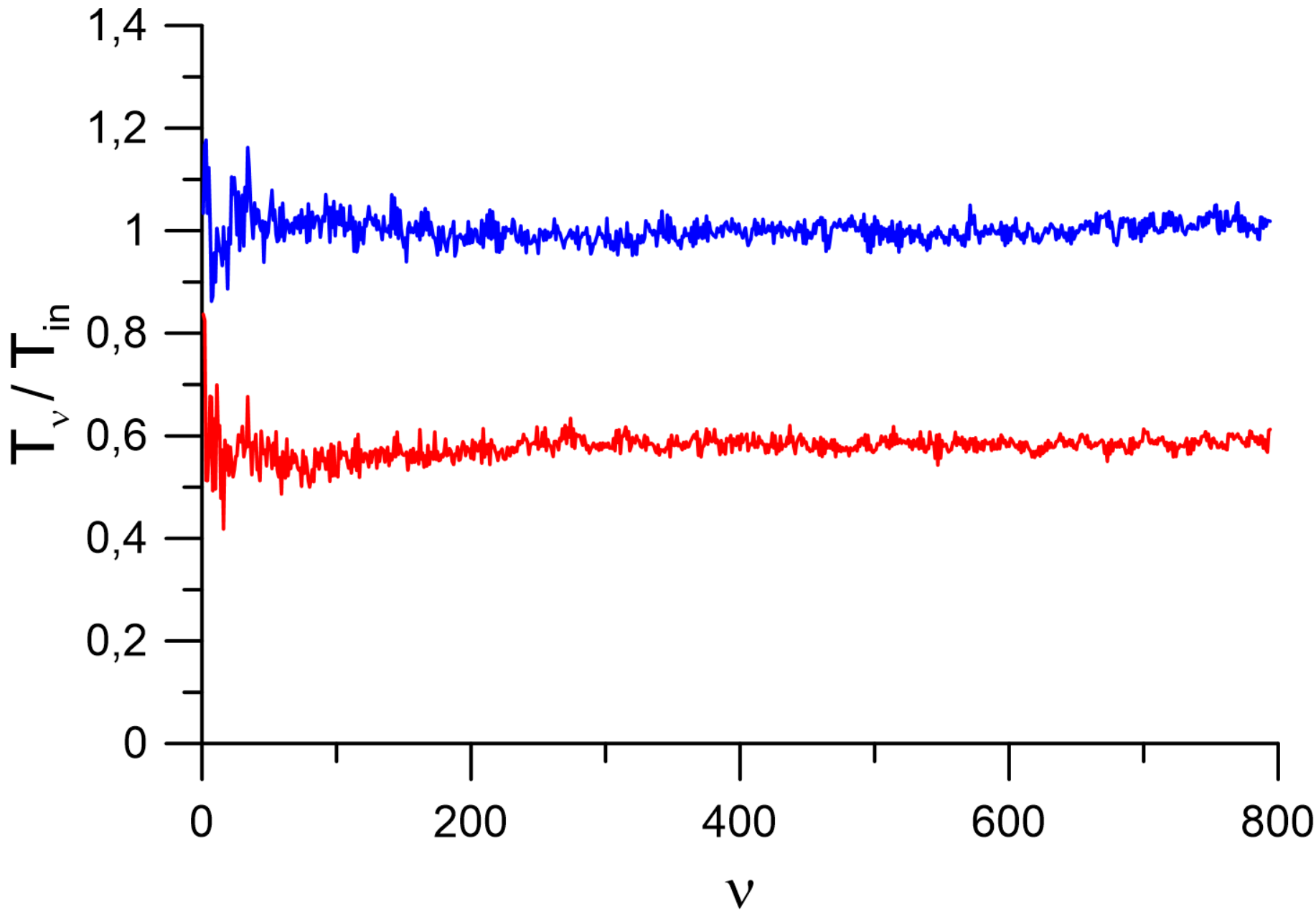}
		\caption{Fraction $T_\nu/T_{in}$ for $T_{in} =50$~nK (upper) and $T_{in} =200$~nK (lower).
		We lower curve corresponds to the case when the initial field was zero outside the region $|x|> x_0 $
		where as the upper one is given for initial field being nonzero everywhere. We observe that the upper curve
		oscillates around unity where as the lower one aroud $0.6$.}
		\label{fig:Tv_Tin}
\end{figure}

In the numerical code for the inhomogeneous case we took $M=1024$ points with the box size  $L=1.9$mm.
Having $g(x)$ and $\triangle_d$ operator we numerically diagonalize Bogoliubov--de Gennes equations (\ref{bog2}).
In one-dimensional case $f_\nu^\pm$ can be chosen real.
Then using the above scheme we construct the initial state for two temperatures $T_{in}=50,200$~nK.
We take $x_0 =  0.95 R$ to be as close the boarder of the $n(x)$ as possible 
in the same time satisfying $\delta n /n \ll 1$. 
As in the homogeneous case we evolve such constructed state according to  (\ref{GP}) taking $g(x,t)= g(x)$ 
allowing system to thermalize to final temperature.
Due to the fact that Bogoliubov method is valid only in the bulk region the 
procedure of extracting temperature from a given state is more complicated than
in the homogeneous case. The bulk region seems to be the most convenient one
to be used in extracting the temperature.
We define 
\begin{equation}\label{Anu}
A_\nu =  \frac{1}{2}\sum_x \Delta x \, f_\nu^+(x)  \frac{\delta  n(x,t)}{\sqrt{n(x)}}. 
\end{equation}
From (\ref{dnt}) we obtain that
\begin{eqnarray*}
A_\nu  =  \sum_{\nu'} c_{\nu \nu'} \mbox{Re}(\alpha_{\nu'})
\end{eqnarray*}
where
\begin{equation}\label{c}
c_{\nu,\nu'} = \sum_x \Delta x \, f_\nu^+(x)f_{\nu'}^-(x). 
\end{equation}
Averaging the square of the above over realizations we obtain
\begin{eqnarray*}
\langle A_\nu^2 \rangle = \sum_{\nu'} c_{\nu \nu'}^2 \langle \mbox{Re}^2(\alpha_\nu') \rangle 
= \sum_{\nu'} c_{\nu \nu'}^2\frac{k_BT}{2\hbar \omega_{\nu'}}  
\end{eqnarray*}
where we used 
\begin{eqnarray*}
\langle \mbox{Re}(\alpha_\nu) \mbox{Re}(\alpha_{\nu'}) \rangle = \delta_{\nu\nu'} \frac{k_BT}{2\hbar \omega_\nu}
\end{eqnarray*}
 implied by the probability distribution (\ref{Pcl}).
The above equations let us to calculate two quantities $ \langle A_\nu^2 \rangle  $
and
\begin{eqnarray*}
 B_\nu = \sum_{\nu'} c_{\nu \nu'}^2\frac{k_B}{2\hbar \omega_{\nu'}}. 
\end{eqnarray*}
In Fig.~(\ref{fig:bv}) we plot $B_\nu$. For modes between 600 and 800 the values
of $B_{\nu}$  oscillate between zero and unity and for modes above 800 the function
$B_\nu$ is almost zero.
The observed very small values of $B_\nu$ are caused by the 
fact that for certain $\nu$, the function $f^+_\nu$ vanishes in the bulk region of the quasicondensate
and is located in the thermal cloud.
In Fig.~(\ref{fig:Tv_Tin}) we plot the values of  $T_\nu/T_{in}$ where 
$ T_\nu = \langle A_\nu^2 \rangle/ B_\nu $ for those $\nu$ for which
$B_\nu$ takes nonzero values. 
Additionally we plot $T_\nu/T_{in}$ obtained in the same way but for 
an initial state for which we set the value of $\psi(x,0)=0$ for $|x|\geq x_0$.
Two plots correspond to $T_{in} = 50,200$~nK.  
We observe that $T_\nu/T_{in}$ is close unity for both temperatures, when we 
took $\psi$ being nonzero in the whole system, while $T_\nu/T_{in}$ is about $0.6$
for $\psi$ being zero at the borders of the system.
We draw the following conclusions from the results presented in figure.
First, the presented results show the necessity of correct introduction of the classical field
outside the bulk region.  For small temperatures 
fraction of the norm of $\psi$ located in that region is very small and 
it is tempting to neglect it. Still in this small fraction a lot of energy of the system
is stored and that is why it cannot be neglected.
Second, the fact $T_\nu \simeq T_{in}$ implies that the temperature of the thermalized state
is  $T_{in}$ with relatively small error. 
Third conclusion is connected with the choice of the field $\psi(x,0)$ outside
the bulk region. It is rather obvious that the thermalized field differs from our choice.
However, the initial field serves only as a state that needs to be thermalized
and any way of constructing the tails of the field is correct as long as 
 $T_\nu \simeq T_{in}$. The fact that this condition is satisfied in the discussed examples 
justifies our choice of the field.

As it can be seen from Eq.~(\ref{Anu}) we diagnosed the temperature  using only density fluctuations 
but
not  phase fluctuations. The reason is that the contribution of each of the modes 
to the density fluctuation does not depend on the mode crucially while in
the phase fluctuations only low modes contribute. This makes it almost impossible
to extract high mode contribution making it useless for temperature diagnostics.

\section{Atomic pairs creation process}\label{sec3}

After discussing the way of preparing the initial state we move to atomic pairs creation process.
In the experiment \cite{Paryz1} the time variation of effective one-dimensional interaction parameter 
is obtained via change of the trap frequencies, which is due to the temporal change of laser intensity.
In the experiment the trap frequencies oscillation is given by
\begin{equation}
\omega_{x,r}(t) = \omega_{x,r} (1+ h \sin \omega_m t) 
\end{equation} 
where $h=0.05$ and $\omega_m = 2\pi \times 2170$ Hz . 
According to \cite{Castin} the relative change of the width $\lambda_{x,r}(t)$ 
is given by equations
\begin{eqnarray}\label{rowew}
\ddot \lambda_r &=& \frac{\omega_r^2(0)}{\lambda_x \lambda_r^3} - \omega_r^2(t) \lambda_r
\\
\ddot \lambda_x &=& \frac{\omega_x^2(0)}{\lambda_x^2 \lambda_r^2} - \omega_x^2(t) \lambda_x.
\end{eqnarray}
As $h\ll 1$ we approximate $\omega_{x,r}^2(t) \simeq \omega_{x,r}^2(0)(1+2h\cos \omega_m t)$. 
We substitute $\lambda_{x,r} = 1 + \delta \lambda_{x,r}$ and linearise the above equations with respect to $\lambda_{x,r}$.
Than the solution is
\begin{equation}\label{dl}
\delta \lambda_r \simeq \frac{h \omega_r}{(2\omega_r)^2 - \omega_m^2} \left(  \omega_m \sin (2 \omega_r t) - 2 \omega_r \sin(\omega_mt) \right).
\end{equation}
The amplitude of $\delta \lambda_x$ is by the factor $\omega_x^2/\omega_r^2 \ll 1$ smaller than the amplitude
$\delta \lambda_r$ and therefore can be neglected. 
The relative change of the radial width leads to the temporal change of $g(x)$ given by the expression
\begin{equation}\label{gtnowe}
g(x,t) \simeq \frac{g(x)}{\lambda_r^2(t)} \simeq g(x) \left( 1 - 2 \delta \lambda_r(t) \right).
\end{equation}
We see that $g(x,t)$ oscillates with two different frequencies  $\omega_m$ and  $2\omega_r$ which are close to each other.

\subsection{Inhomogeneous system}

Now we proceed with numerical simulations towards the true experimental situation.
We take the initial state as the final  state obtained in the previous Section
and evolve it for $t_f = 25$~ms  using GP equation (\ref{GP}) with $g(x,t)$ given by Eqs.~(\ref{dl}) and (\ref{gtnowe}).
\begin{figure}[htb]
	\centering
		\includegraphics[width=0.48\textwidth]{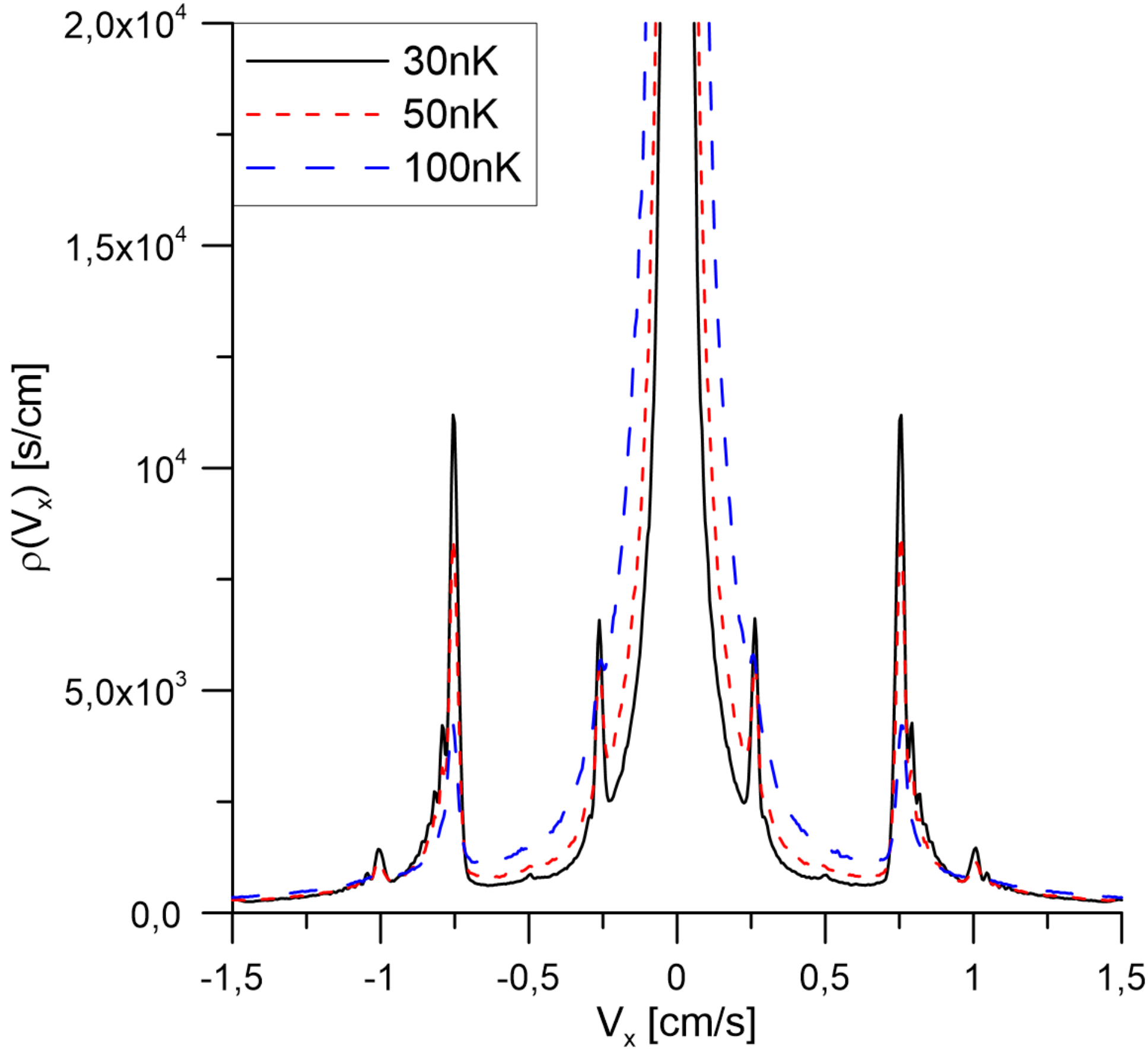}\\
		\includegraphics[width=0.48\textwidth]{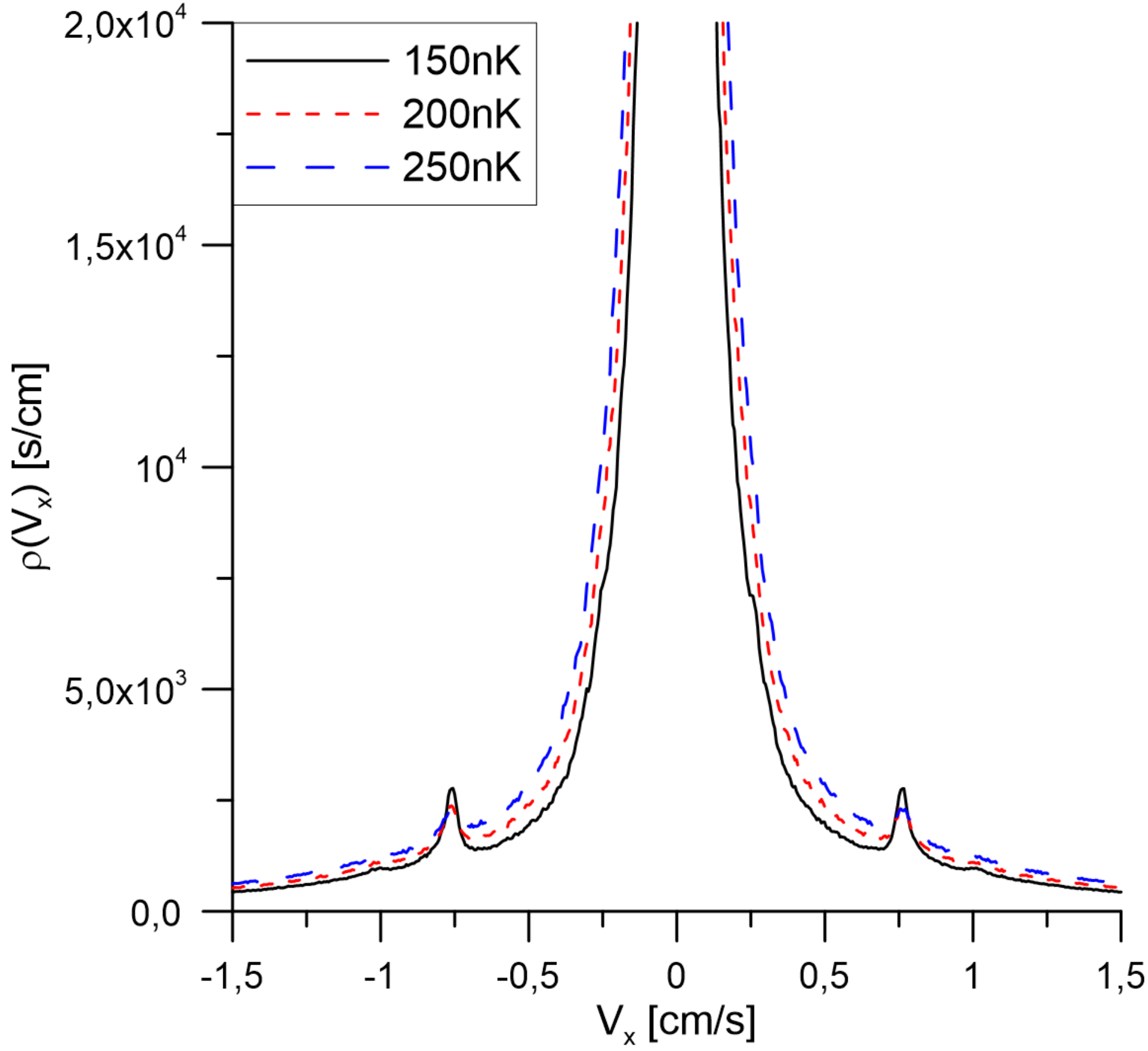}
		\caption{Momentum density $\rho(v_x) = \frac{1}{N_r} \sum_{r=1}^{N_r} |\psi_r(v_x,t_f)|^2 $  averaged over $N_r = 5000$ realizations. }
		\label{fig:momh}
\end{figure}
We repeat the procedure many times to obtain the average over realizations.
In Fig.~(\ref{fig:momh}) we plot the momentum density of the classical field
$\rho(V_x) = \frac{1}{N_r} \sum_{r=1}^{N_r} |\psi_r(V_x,t_f)|^2 $  averaged over $N_r = 5000$ realizations
for a few initial temperatures.  
The field is normalized  $\sum_{v_x} \Delta v \, |\psi(v_x,t_f)|^2 = N$ with $\Delta v = \frac{2\pi \hbar}{mL}$. 
We notice huge central peak which is given by the quasicondensate distribution
which width grows with temperature as expected.
In the tails of the distribution we notice pairs of peaks. 
For low temperatures (first panel) we notice three pairs of peaks with velocities $|V_x| \simeq 0.25, 0.8, 1$~cm/s.
Using the homogeneous case calculation presented in the following subsection we find that 
the peaks with velocities $0.8$~cm/s and $1.0$~cm/s correspond to frequencies $\omega_m$ and $2 \omega_r$ respectively.
The third peak with velocity $0.25$~cm/s corresponds roughly to frequency $2 \omega_r - \omega_m$ which is
a clear sign of frequency mixing taking place in the system.
The highs of all the peaks produced by the temporal modulation of the interaction parameter
decrease with increasing temperature so that at higher temperatures (second panel) only the 
$\omega_m$ peaks are visible.

In the experiment  the system is suddenly released from the trap at time $t_f$.
We simulate this part of the experiment to check if  expansion changes the momentum density.
To model the ballistic expansion we solve Eq.(\ref{rowew}) obtaining in that case
$\lambda_r(t) = \sqrt{1+\omega_r^2 t^2}$. It gives us effective change
of the one-dimensional interaction parameter $g(x,t) = g(x)/(1+\omega_r^2t^2)$.
With such interaction parameter and taking $V(x)=0$ to model the expansion we
 further evolved the GP equation (\ref{GP}) for time $t_0 \simeq 30/\omega_r$.
After that time $g(t)$ is so small that it can be  practically neglected. 
Then the evolution is linear with  unchanged momentum distribution.
We calculated the final momentum density and compared it with the one before the release of the trap 
(at the end of modulation period). We have not found any noticeable differences. 

Let us now compare the momentum density for $T=200$~nK (which is the temperature in the experiment \cite{Paryz1}) 
plotted in Fig.~(\ref{fig:momh}) with the one measured in the experiment that can be seen in Fig.~1e of \cite{Paryz1}.
We notice that the position of the peaks in the experiment resembles the one 
coming from numerical simulations. The same applies to the amplitude of the peak with respect
to the background given by the tails of the central peak.
Thus we notice a good agreement between the theoretical calculation and the experimental results.

In \cite{Paryz1} the authors  compared also intensity differences in the two peaks.
One of the possibilities to do this is to calculate
so-called number squeezing parameter defined as
\begin{equation}\label{sdef}
s(t) = \frac{\left\langle \left(\hat N_+(t) - \hat N_-(t) \right)^2 \right\rangle - \left\langle \left(\hat N_+(t) - \hat N_-(t) \right)\right\rangle^2
  }{\left\langle \left(\hat N_+(t)  + \hat N_-(t) \right)\right\rangle}
\end{equation}
where 
\begin{equation}\label{Npmdef}
\hat N_\pm(t) = \sum_{|k \mp k_0| < \Delta k_0} \Delta k \, \hat \Psi^\dagger(k,t) \hat \Psi(k,t)
\end{equation}
is the operator describing number of particles in the peak which maximum is located at $\pm k_0$.
The condition  $s<1$ together with $\langle \hat N_+ \rangle = \langle \hat N_- \rangle$ implies the
particle entanglement of the quantum state \cite{jan}.
In the classical field method the above takes the form
\begin{eqnarray*}
N_{\pm}(t_f) = \sum_{|v \mp v_0| < \Delta v_0} \Delta v \, |\psi(v_x,t_f)|^2
\end{eqnarray*}
and
\begin{equation}\label{scl}
s(t) = \frac{\langle \left( N_+(t) -  N_-(t) \right)^2 \rangle - \langle \left( N_+(t) -  N_-(t) \right)\rangle^2  
}{\langle \left( N_+(t)  +  N_-(t) \right)\rangle }.
\end{equation}
where $\Delta v = \frac{2\pi \hbar}{mL}$  and $\psi(v_x,t_f)$ is the classical field in single realization at final time $t_f$.
Using the above formulas we calculated numerically $s(t_f)$ for experimental temperature $T=200$~nK and for a few values of $\Delta v_0$.
We found the minimal value of $s(t_f)$ being close to $9$.
This is in agreement with experimental measurements as the authors of \cite{Paryz1} report 
lack of sub-Poissonian fluctuations \cite{dopiska2}.

Finally we comment on number squeezing calculation in the classical field method.
Here we face the problem of operator ordering.
We note that in the formula for the number squeezing parameter the 
numerator is quadratic in the number operator.
Thus the difference coming from different ordering of the operators in the numerator
is equal roughly to the number operator which is present in the denominator 
of the number squeezing formula.
This shows that different choices of operator ordering result in 
difference in the number squeezing parameter of the order of unity.
Thus the uncertainty of $s$ parameter calculation given by the classical field method can be estimated to be equal to one.
So in fact the result in the inhomogeneous case is $s \simeq 9 \pm 1$ which
still gives lack of sub-Poissonian fluctuations in the system.

We briefly summarize that the classical field simulation results are in agreement with the experimental measurments.
However, to understand the theoretical results we move to homogeneous case analysis.

\subsection{Homogeneous case}

To simplify the analysis we take 
\begin{eqnarray} \nonumber
g(t) &=& g(0) \left(1 + 2  \frac{h \omega_r}{(2\omega_r)^2 - \omega_m^2}  2 \omega_r \sin(\omega_mt) \right)
\\ \label{gth}
&=& g(1+\epsilon \sin \omega_m t).
\end{eqnarray}
which is equal to $g(x=0,t)$ given by (\ref{gtnowe}) with the part $\sin (2 \omega_r t)$ neglected. 
As before we chose the density to be given by the maximal density of the homogeneous case.

\subsubsection{Bogoliubov method}\label{Bog0}

At first we perform Bogoliubov analysis of the pair production process similar to that in \cite{nasza}.
In such case we do not use classical field approximation but stay at quantum field theory analysis.
In the Bogoliubov method we use the density and phase operator representation of the field operator
\cite{gora}
\begin{eqnarray*}
\hat \Psi = e^{i ( \phi +\hat \phi)} \sqrt{n + \delta \hat n}. 
\end{eqnarray*}
together with the mode decomposition
\begin{eqnarray*}
\delta \hat n(x,t) &=& \sqrt{n} \sum_k f_k^-(x) \hat a_k(t) + \mbox{h.c.}
\\
\delta \hat \phi(x,t) &=& \frac{1}{2\sqrt{n}} \sum_k -i f_k^+(x) \hat a_k(t) + \mbox{h.c.}.
\end{eqnarray*}

The equation of motion for the density $\delta \hat n$, phase $\phi$ and  phase operator $\hat \phi$  
reads \cite{gora}  
\begin{eqnarray*} \nonumber
 - \hbar \frac{\partial \phi}{\partial t}  &=& g(t) n 
\\ 
 \hbar (\partial \delta \hat n/\sqrt{n})/\partial t &=& - \frac{\hbar^2}{2m} \frac{\partial^2}{\partial x^2} ( 2 \hat \phi \sqrt{n} )
\\
 - \hbar \frac{\partial (2\sqrt{n} \hat \phi)}{ \partial t} &=& \left( - \frac{\hbar^2}{2m} \frac{\partial^2}{\partial x^2}
+ 2 g(t) n \right) \frac{\delta \hat n}{\sqrt{n}}.
\end{eqnarray*}
Inserting the mode decomposition into the above we find
\begin{eqnarray*}
\frac{\mbox{d}}{\mbox{d} t} \hat a_k = - i \omega_k \left( \frac{E_k + (g(t) + g)n}{E_k + 2 gn} \hat a_k + 
\frac{(g(t) - g)n}{E_k + 2 gn} \hat a_{-k}^\dagger  \right).
\end{eqnarray*}
Here functions $f_k^\pm$ are given by Eq. (\ref{fhom}).
We now take $g(t) = g(1+\epsilon \sin \omega_m t)$ given by Eq.~(\ref{gth})
where $\epsilon \ll 1$ and approximate
 $\frac{E_k + (g(t) + g)n}{E_k + 2 gn} \simeq 1 $. We additionally assume that
$|\omega - \omega_k| \ll \omega$ which enables us to use rotating wave approximation so that the above equation takes the form
\begin{eqnarray*}
i \frac{\mbox{d}}{\mbox{d} t} \hat a_k =  \omega_k \left(  \hat a_k +
  i\frac{\epsilon gn}{2 (E_k + 2 gn)}  e^{-i \omega_m t} \hat a_{-k}^\dagger  \right),
\end{eqnarray*}
with the solution 
\begin{eqnarray} \nonumber
\hat a_k(t) &=& \left( \cosh \Omega_k t + i \frac{\Delta_k}{\Omega_k} \sinh \Omega_k t \right) e^{-i\omega_m t/2} \hat a_k(0)
\\ \label{akt1}
& & + \frac{\delta_k}{\Omega_k} \sinh \Omega_k t e^{-i\omega_m t/2} \hat a_{-k}^\dagger(0)
\end{eqnarray}
where 
\begin{eqnarray*}
\delta_k &=& \omega_k \frac{\epsilon gn}{2 (E_k + 2 gn)} 
\\ 
\Delta_k &=& \omega_m/2 - \omega_k 
\ \ \ \
\Omega_k = \sqrt{\delta_k^2 - \Delta_k^2}.
\end{eqnarray*}
We find the resonance at  $k_0$ satisfying  $\omega_m = 2\omega_{k_0}$ with 
the resonance width approximately equal to  $\delta_{k_0}$.

The above describes quasiparticle properties. We now turn our attention
to the particle properties. In Appendix \ref{ap1} we establish an approximate
connection between quasiparticle and particle properties of the system.
We find that the particle momentum density is approximately equal to
\begin{eqnarray} \nonumber
\langle \hat \Psi^\dagger (k,t) \hat \Psi(k,t) \rangle &=& N \rho(k) + \rho(k-k_0) 
\langle \hat b^\dagger_{k_0}(t) \hat b_{k_0}(t) \rangle
\\ \label{dens00}
& & + \rho(k+k_0) \langle \hat b^\dagger_{-k_0}(t) \hat b_{-k_0}(t) \rangle
\end{eqnarray}
where
\begin{equation}\label{bk10}
\hat b_k(t) = u_k \hat a_k(t) - v_k \hat a_{-k}^\dagger(t)
\end{equation}
and
\begin{eqnarray*}
\sum_k \Delta k \, \rho(k) = 1
\end{eqnarray*}
 with $\Delta k = \frac{2\pi}{L}$.
By inspecting Eq.~(\ref{dens00})
we find that the particle momentum distribution has three peaks.
Central one equal to $N\rho(k)$ represents the quasicondensate momentum distribution.
Two other peaks represent the  resonant quasiparticle modes
centered around $k = \pm k_0$.
We notice that the shape of the peaks is given by the quasicondensate momentum 
distribution. 
Substituting experimental values we find $ \hbar k_0/m \simeq 0.8 \, \mbox{cm}/\mbox{s}$ which is the same 
as the value of the center of the peaks in the momentum distribution observed experimentally.

In the experiment \cite{Paryz1} the authors measured the properties of the
number of particles in both peaks.  Above we introduced operator $\hat N_{\pm}$ defined by
Eq.~(\ref{Npmdef})
describing number of particles in the peak located at $\pm k_0$.
Assuming that $\Delta k_0$ present in the definition of $\hat N_\pm$ is such that it covers most of the peak
 we show in Appendix \ref{ap1} that
\begin{equation} \label{Npm0}
\hat N_\pm(t) \simeq \hat b_{\pm k_0}^\dagger(t) \hat b_{\pm k_0}(t).
\end{equation}
Applying Bogoliubov results given by Eq.~(\ref{akt1}) and using
Eq.~(\ref{bk10}) we find 
\begin{equation}\label{Npm1}
\langle \hat N_\pm(t) \rangle \simeq  \langle \hat b_{\pm k_0}^\dagger(t) \hat b_{\pm k_0}(t)
\rangle
= u_{k_0}^2 n_{k_0}(t) + v_{k_0}^2(n_{k_0}(t)+1) 
\end{equation}
where
\begin{eqnarray} \nonumber
n_{k_0}(t) = \langle \hat a_{k_0}^\dagger(t) \hat a_{k_0}(t) \rangle &=& n_{k_0} \cosh^2 \delta_{k_0} t 
\\ \label{nkt}
& &
+ (n_{k_0}+1) \sinh^2 \delta_{k_0} t 
\end{eqnarray}
and we took initial state as a thermal one with
\begin{equation}\label{nkk}
n_k = \langle \hat a_k^\dagger (0) \hat a_k(0) \rangle  
= \left( \exp \left( \hbar \omega_k/ k_BT  \right) -1\right)^{-1}.
\end{equation}
Taking the experimental parameters we find
$\delta_{k_0} \simeq 170$~Hz, 
 $v_{k_0}^2 = u_{k_0}^2-1 \simeq 0.55$ and 
$n_{k_0} \simeq 3.36$ for $T=200$~nK and $t=t_f=25$ms.
For such values we find $\langle N_\pm(t_f) \rangle \simeq 2\times 10^4 $ which is much larger than the 
value measured in the experiment. 
We also calculate number squeezing parameter defined by Eq.~(\ref{sdef}).
From Eqs.~(\ref{bk10}) and (\ref{Npm0})  we find
\begin{equation}\label{npm11}
\left(\hat N_+ - \hat N_- \right)(t) = \left( \hat a_{k_0}^\dagger \hat a_{k_0} -  \hat a_{-k_0}^\dagger \hat a_{-k_0}\right)(t)
\end{equation}
where we used the normalization condition $u_k^2 - v_k^2 =1$. 
Moreover from Eq.~(\ref{akt1}) we find that
\begin{eqnarray*}
\left(\hat a_k^\dagger \hat a_k  -  \hat a_{-k}^\dagger \hat a_{-k} \right) (t)
 = \left(\hat a_k^\dagger \hat a_k  -  \hat a_{-k}^\dagger \hat a_{-k} \right) (0).
\end{eqnarray*}
As a result the numerator of the formula for the squeezing parameter is constant in time
and equals to
\begin{eqnarray*}
&& \left\langle \left(\hat N_+(t) - \hat N_-(t)\right)^2 \right\rangle - \left\langle \left(\hat N_+(t) - \hat N_-(t)  \right)\right\rangle^2  
\\
&& = 2 n_{k_0}( n_{k_0} + 1).
\end{eqnarray*}
The denominator of the mentioned formula is equal to $\langle \left(\hat N_+(t)  + \hat N_-(t) \right)\rangle$
and according to Eqs.~(\ref{Npm1}) and (\ref{nkt}) grows in time.
For the experimental parameters we find that $s(t_f) \simeq 7 \times 10^{-4}$ which is much smaller than unity
which shows disagreement with experimental results where $s(t_f)$ is above unity.

As we see, the above calculations of peak population properties show huge discrepancies
with experimental measurements. This shows the necessity of taking into account
the interaction between quasiparticles neglected in the Bogoliubov method.
To take into account them we move to classical field method.

\subsubsection{Classical field analysis}

\begin{figure}[htb]
  \includegraphics[width=0.48\textwidth]{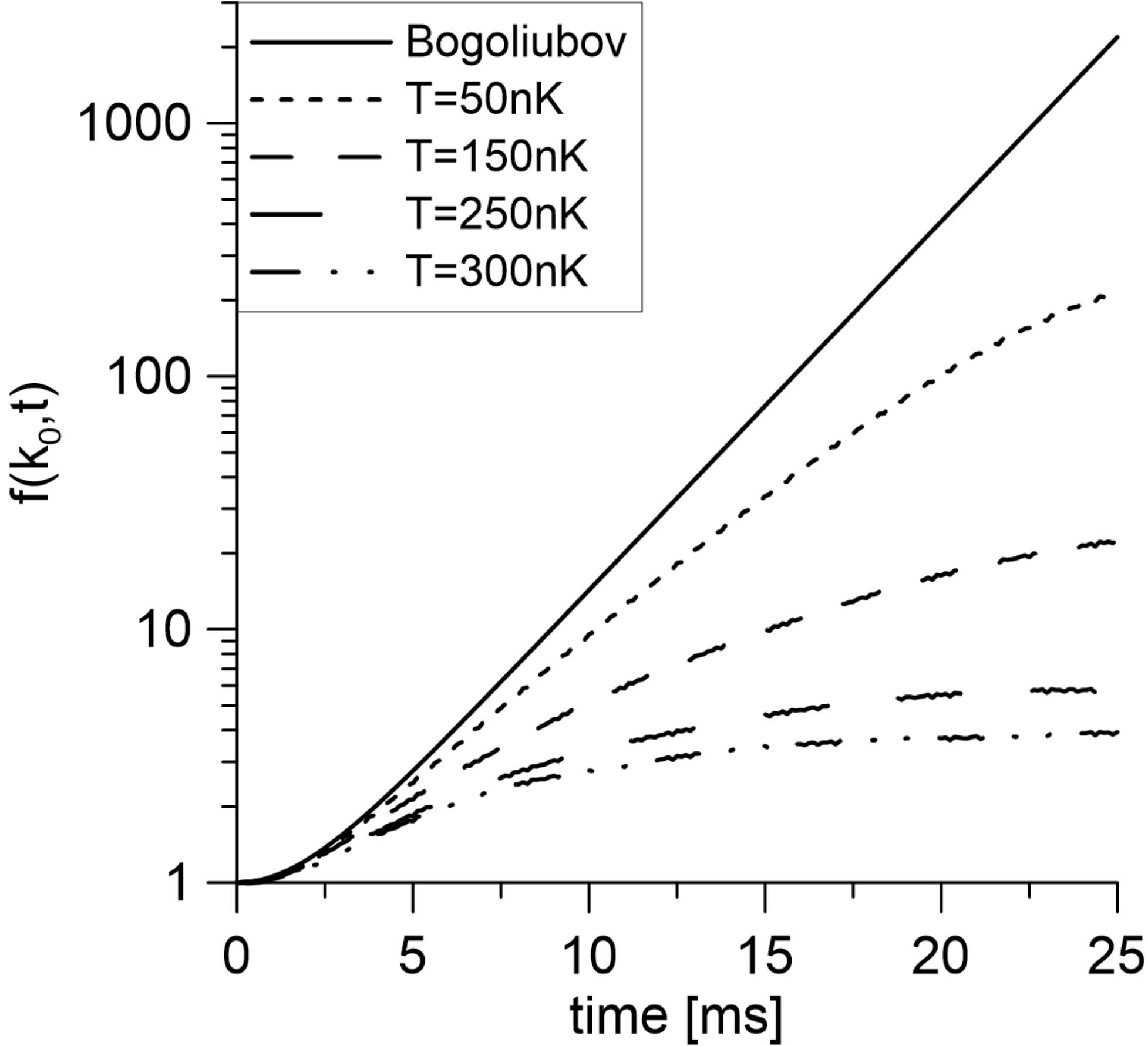}
  \caption{The quasiparticle relative mode occupation $f(k_0,t) = \langle |\alpha_{k_0}(t)|^2 \rangle/ \langle |\alpha_{k_0}(0)|^2 \rangle $
	for the mode in resonance as a function of time for few temperatures.}
  \label{wj}
\end{figure}

The method is defined as follows.
We take the initial state $\psi(x,0)$ as the thermal state obtained in the previous Section.
Then we evolve the GP equation (\ref{GP}) for $t_f = 25$~ms as in the experiment
taking $V(x)=0$ and $g(x,t) = g(t)$ given by Eq.~(\ref{gth}).
We repeat the procedure many times to obtain the average over realizations. 
First we calculate $f(k_0,t) = \langle |\alpha_{k_0}(t)|^2 \rangle/ \langle |\alpha_{k_0}(0)|^2 \rangle $
where the quasiparticle amplitude $\alpha_{k_0}(t)$ is given by Eq.~(\ref{alphat}).
In Fig.~(\ref{wj}) we plot this quantity for various temperatures.  
For comparison we draw the Bogoliubov prediction in the classical field method derived in Appendix \ref{ap2} 
\begin{eqnarray*}
f(k_0,t) = \frac{\langle |\alpha_{k_0}(t)|^2 \rangle}{ \langle |\alpha_{k_0}(0)|^2 \rangle} =  \cosh (2 \delta_{k_0} t).
\end{eqnarray*}
We notice dramatic drop of production of quasiparticles with the increase of the temperature.
For low temperatures the productions tends to Bogoliubov result.
We also note that for smaller temperatures the quasiparticles are constantly
produced in time, whereas for higher temperatures the number of quasiparticles produced
tends to a constant, practically stopping after some critical time.

We interpret this facts in the spirit of the papers \cite{nasza,busch}.
In \cite{nasza} three dimensional analogue of the system considered here was discusses.
There the interaction between quasiparticles was taken into acount in the description of the
parametric amplification process. In the approximation assuming a lack of memory of
self energy functions,   
interaction between quasiparticles enters the process only through single parameter
$\gamma_k$ which is simply the inverse of quasiparticle lifetime.
It was found that the pair production process practically stops in time when 
$\gamma_{k_0}$ gets larger than the amplification parameter $\delta_{k_0}$
defined in the same way as in the Bogoliubov method described above.
Interesting idea of describing the systems similar to the considered one
was presented in \cite{busch}.
There the authors consider Bogoliubov model as presented above.
However, the interaction between quasiparticles is substituted by interaction
of quasiparticles with large reservoir. 
The authors assume that the response of the reservoir to external perturbation is
instantaneous in time 
(which is equivalent to the above mentioned assumption of lack of memory of self energy functions used in \cite{nasza}). 
Then it is not surprising that  the interaction with the reservoir 
is effectively described by $\gamma_k$ and that the results of \cite{busch} are the same as the one obtained in \cite{nasza}.
Still the description presented in \cite{busch} that is quite general, enables to use the results
obtained in \cite{nasza,busch} for the one-dimensional system considered here.
They could be used in a straightforward way if the assumption of lack of memory is satisfied.
This can be verified by looking at the quasiparticle decay curve which is then of exponential type.
Unfortunately, as we shall see below, this is not the case in our system.
However, in the model considered in \cite{busch} or alternatively in the models
of two modes coupled to a reservoir widely used in quantum optics \cite{ksiazka},
the quasiparticle decay curve can take different shapes depending on the 
reservoir memory functions used.
Thus we expect that the condition $\delta_{k_0} \simeq \gamma_{k_0}$
for  stopping of pair production process still applies to our system
with $\gamma_{k_0}$ describing the width of the quasiparticle decay curve.

\begin{figure}[htb]
  \includegraphics[width=0.48\textwidth]{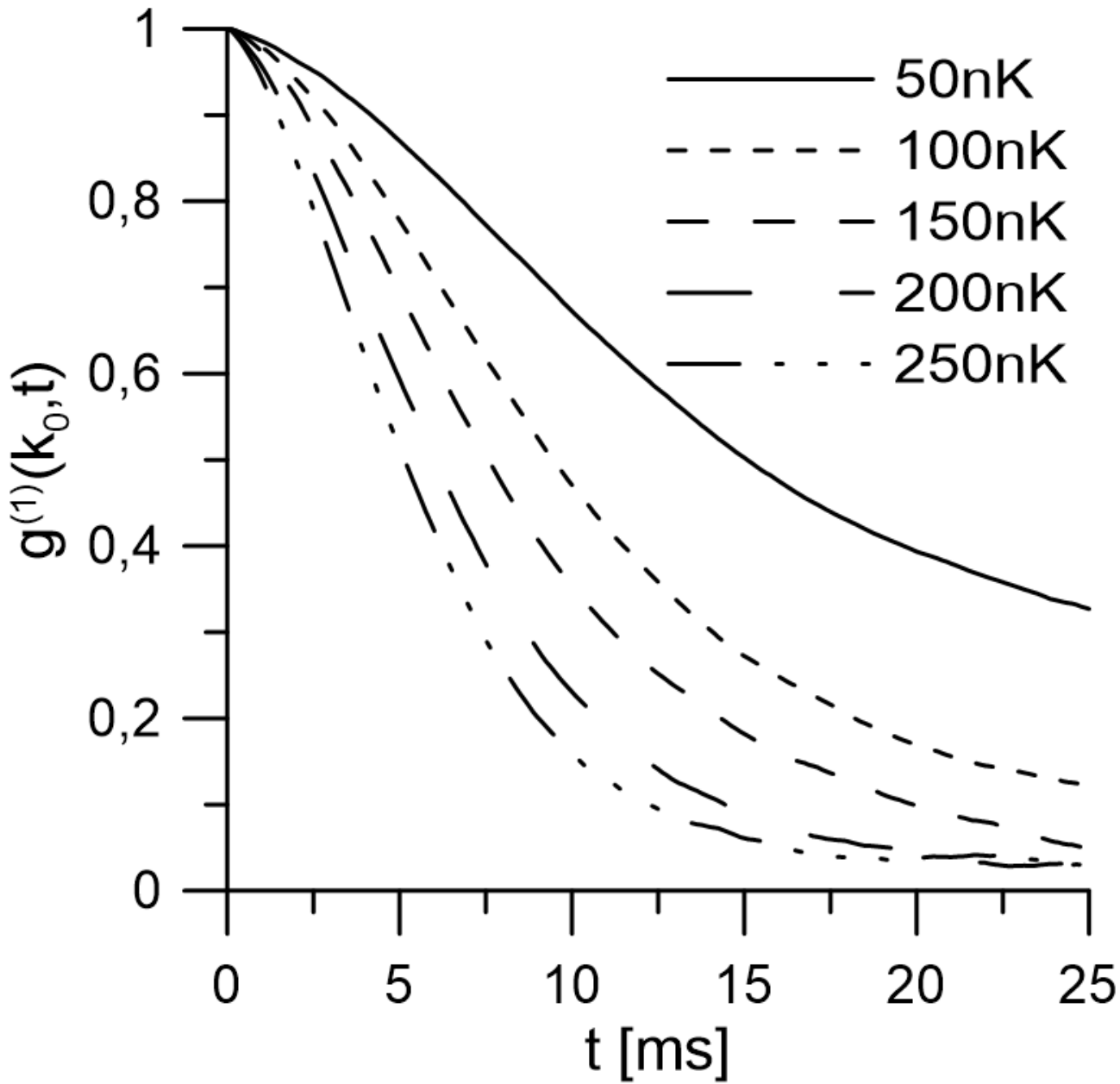}
  \caption{The  normalized single particle correlation function $g_k(t)$ for the resonance mode as 
as a function of time for few temperatures.}
  \label{gj}
\end{figure}

Below we check  if the condition described above applies to our system.
We calculate numerically the normalized single particle correlation function 
\begin{eqnarray*}
g^{(1)}(k_0,t) = \frac{|\langle \alpha_{k_0}^*(0) \alpha_{k_0}(t) \rangle|}{\langle |\alpha_{k_0}(0)|^2 \rangle }
\end{eqnarray*}
for the resonant mode.
The result of numerical calculation are presented in Fig.~(\ref{gj}).
Looking at this figure we clearly see that the curves are not of exponential type.
Still the halfwidth decreases with increasing temperature as expected.
Looking at Fig.~(\ref{wj}) we find that the quasiparticle production process practically stops
for temperature around $150$nK.
For that temperature we find $\gamma_{k_0} \simeq 100$Hz which is close to $\delta_{k_0}=170$Hz calculated above.
It shows that the derived condition applies.

Additionally we calculate the squeezing parameter.
In order to do it we rewrite the Eqs.~(\ref{bk10}) and  (\ref{Npm1})
substituting $\hat b_k \rightarrow \beta_k$. We obtain
\begin{equation}\label{bkcl}
\beta_k(t) = u_k \alpha_k(t) - v_k \alpha_{-k}^*(t).
\end{equation}
and
\begin{equation}\label{Npmcl}
N_\pm(t) = |\beta_{\pm k_0}(t)|^2.
\end{equation}
Using Eqs.~(\ref{bkcl}) and (\ref{Npmcl}) where $\alpha_k(t)$ is given by Eq.~(\ref{alphat})
we calculate all the observables needed in the calculation of the squeezing parameter
given by Eq.~(\ref{scl}).
For the temperature  $T=200$nK we find $s(t_f) \simeq 3.3$.
This is larger than unity but smaller than the value obtained in the inhomogeneous system case
being close to $9$.
We think that the reason for the observed difference is the following.
Looking at velocity distribution shown at Fig.~\ref{fig:momh} we notice
that for temperature $200$~nK the peak height compared to background density coming
from the quasicondensate distribution is smaller than unity.
It means that more than half of particles
contributing to the 
 number of particles in the peak 
comes from thermal distribution. This particles are not correlated in velocities and 
rather contribute to the increase of the number squeezing parameter.
On the other hand the value in the homogeneous case is obtained
assuming lack of quasicondensate particles in the side peaks.
That is probably why the value of $s(t_f)$ calculated in the homogeneous case is smaller than in the
non-homogeneous case. 
However, it is important to notice that both values give lack of sub-Poissonian fluctuations in the system.

\section{Summary}\label{sec4}

In the present paper we performed numerical simulations of  
experiment \cite{Paryz1} using classical field method. 
The atom pairs were created using temporal modification of the effective interaction parameter.
The results found are in full agreement with the experimental measurements.
To understand the theoretical results we additionally 
analyzed the homogeneous analogue of the experimental system. Analytical calculations within 
Bogoliubov method were performed together with numerical simulations using classical field approximation. 
The results of the Bogoliubov method gave pair production being much larger than observed experimentally. 
The classical field analysis showed agreement with Bogoliubov results for temperatures 
significantly smaller than in the experiment.
For higher temperatures it predicted dramatic decrease of number of pairs produced
being in agreement with the experiment.
We additionally calculated the number squeezing parameter for both homogeneous and
inhomogeneous system. We found that number squeezing does not take place
which is in agreement with experimental measurements.
We interpreted this results in the spirit of findings presented in \cite{nasza,busch}.
There it was shown that the interaction between quasiparticles omitted in the Bogoliubov method
(and accounted for in the classical field approximation) may dramatically influence the pair production process
as well as the value of number squeezing parameter.
It was shown that the pair production process practically stops when the parameter $\delta_{k_0}$
describing enhancement of the resonant mode, derived within Bogoliubov method, is roughly equal
to the quasiparticle decay constant $\gamma_{k_0}$ which appears as a consequence of interaction between quasiparticles.
We numerically calculated $\gamma_{k_0}$ and atom pair production as a function of temperature together with
$\delta_{k_0}$ parameter.
We have shown that indeed the pair production process gets frozen when $\gamma_{k_0} \simeq \delta_{k_0}$.
Additionally we have shown that the experiment is in the regime when the pair production process gets frozen
after short time. This explains the fact of relatively small number of generated atom pairs observed in the experiment.

\begin{acknowledgments}
We acknowledge the discussions with Chris Westbrook, Denis Boiron and Dimitri Gangardt.
\end{acknowledgments}

\appendix

\section{Connection between quasiparticle and particle properties for homogeneous system}\label{ap1}

In this Appendix we establish approximate relation between quasiparticle and
particle properties of the homogeneous system.
We have
\begin{eqnarray*}
\hat \Psi = e^{i ( \phi +\hat \phi)} \sqrt{n + \delta \hat n} 
\end{eqnarray*}
together with the  mode decomposition 
\begin{eqnarray*}
\delta \hat n(x,t) &=& \sqrt{n} \sum_k f_k^-(x) \hat a_k(t) + \mbox{h.c.}
\\
\delta \hat \phi(x,t) &=& \frac{1}{2\sqrt{n}} \sum_k -i f_k^+(x) \hat a_k(t) + \mbox{h.c.}
\end{eqnarray*}
 where $f_k^\pm$ are given by Eq.~(\ref{fhom}).
Modes of the system can be divided into low and high energy ones.
Low energy modes are highly populated and responsible for phase fluctuations of the system.
The high energy modes population in the equilibrium state is much smaller than the population of low lying modes.
We divide $\hat \phi = \hat \phi_l + \hat \phi_h$, $\delta \hat n = \delta n_l + \delta n_h$
and approximate
\begin{eqnarray*}
\hat \Psi(x) = e^{i ( \phi +\hat \phi)} \sqrt{n + \delta \hat n}
\simeq e^{i ( \phi +\hat \phi_l)} \sqrt{n} \left( 1 - i \hat \phi_h + \frac{\delta \hat n_h}{2 n} \right).
\end{eqnarray*}
Inserting into the above the mode decomposition one obtains
\begin{eqnarray*}
 \hat \Psi(x) \simeq \frac{1}{\sqrt{L}} e^{i ( \phi +\hat \phi_l)} \left( \sqrt{N} + \sum_{k \in h} u_k e^{ikx} \hat a_k - v_k e^{-ikx} \hat a_k^\dagger \right) 
\end{eqnarray*}
where we used $f_k^\pm = u_k \pm v_k.$
We further simplify the above by treating the phase operator of the low energy modes (which are all highly populated) classically
$\hat \phi_l \rightarrow \phi_l$. Having done that we introduce classical field of the low energy modes defined as
\begin{equation}\label{psic1}
\psi_c(x,t) =  \frac{1}{\sqrt{L}} e^{i (\phi(t) + \phi_l(x,t))}.  
\end{equation}
Using that field and additionally introducing  
\begin{eqnarray*}
\hat b_k(t) = u_k \hat a_k(t) - v_k \hat a_{-k}^\dagger(t)
\end{eqnarray*}
we arrive at
\begin{eqnarray*}
\hat \Psi(x,t) \simeq \psi_c(x,t) \left( \sqrt{N} + \sum_{k \in h}  e^{ikx} \hat b_k(t) \right).
\end{eqnarray*}
Inserting the above into
\begin{eqnarray*}
\hat \Psi(k,t) = \frac{1}{\sqrt{2\pi}} \sum_j \Delta x  \, e^{-ikx} \hat \Psi(x,t)
\end{eqnarray*}
we arrive at
\begin{equation}\label{hatPsik}
\hat \Psi(k,t)  \simeq  \sqrt{N} \psi_c(k,t) + \sum_p \psi_c(k-p,t) \hat b_p(t)
\end{equation}
where
\begin{equation}\label{psic12}
\psi_c(k,t) = \frac{1}{\sqrt{2\pi}} \sum_x \Delta x  \, e^{-ikx}  \psi_c(x,t).
\end{equation}
We assume that the influence of the high energy sector on the dynamics of the low
energy sector can be neglected. Thus the classical phase $\phi(t) +\phi_l(x,t)$ describes the 
evolution of the thermal state. This implies that the averages of the field $\psi_c(x,t)$
are the thermal state averages that do not depend on time.

We further assume that among all high energy modes only two resonance modes
with  $k=\pm k_0$ have significant population.
Thus we have
\begin{eqnarray} \nonumber
\langle \hat \Psi^\dagger (k,t) \hat \Psi(k,t) \rangle &=& N \rho(k) + \rho(k-k_0) 
\langle \hat b^\dagger_{k_0}(t) \hat b_{k_0}(t) \rangle
\\ \label{dens0}
& & + \rho(k+k_0) \langle \hat b^\dagger_{-k_0}(t) \hat b_{-k_0}(t) \rangle
\end{eqnarray}
where
\begin{eqnarray*}
\rho(k) = \langle |\psi_c(k,t)|^2 \rangle. 
\end{eqnarray*}
In the above formula we used the fact that the average $\langle |\psi_c(k,t)|^2 \rangle $
does not depend on time.
Using Eqs.~(\ref{psic1}) and (\ref{psic12}) we find that $\rho$ is normalized to unity i.e.
\begin{eqnarray*}
\sum_k \Delta k \, \rho(k) = 1
\end{eqnarray*}
 where $\Delta k = \frac{2\pi}{L}$. 
We introduce operators
\begin{eqnarray*}
\hat N_\pm = \sum_{|k \mp k_0| < \Delta k_0} \Delta k \, \hat \Psi^\dagger(k) \hat \Psi(k).
\end{eqnarray*}
Using Eq.~(\ref{hatPsik}) and the fact that $\psi_c^*(k)\psi_c(k \pm k_0) \simeq 0$
we find 
\begin{eqnarray*}
\hat N_\pm = \sum_{|k \mp k_0| < \Delta k_0} \Delta k \, |\psi_c(k \mp k_0)|^2 \hat b_{\pm k_0}^\dagger \hat b_{\pm k_0}.
\end{eqnarray*}
From Eqs.~(\ref{psic1}) and (\ref{psic12}) we obtain
\begin{eqnarray*}
\sum_{|k \mp k_0| < \Delta k_0} \Delta k \, |\psi_c(k \mp k_0)|^2 \simeq 1.
\end{eqnarray*}
To obtain the above we assumed that $\Delta k_0$ is such that it covers most of the peak.
As a result we find  
\begin{eqnarray*}
\hat N_\pm \simeq \hat b_{\pm k_0}^\dagger \hat b_{\pm k_0}.
\end{eqnarray*}

\section{The Bogoliubov method in the classical field approximation }\label{ap2}

In this Appendix we use the results of the  Bogoliubov method described in Sec. \ref{Bog0}
 to find their analogue in the classical field approximation. 
In classical field method we substitute the annihilation and creation operators by c-numbers
$\hat a_k \rightarrow \alpha_k$. 
Then the averages over $\alpha_k(0)$ are  calculated using the probability distribution given by Eq.~(\ref{Pcl}).
For example 
\begin{eqnarray*}
\langle |\alpha_k(0)|^2 \rangle = \frac{k_B T}{\hbar \omega_k}
\end{eqnarray*}
This corresponds to quantum average:
\begin{eqnarray*}
n_k =\langle \hat a_k^\dagger (0) \hat a_k(0) \rangle  
= \left( \exp \left( \hbar \omega_k/ k_BT  \right) -1\right)^{-1}
\end{eqnarray*}
in the limit $ n_k \gg 1$. Applying the above procedure to Eq.~(\ref{akt1}) we find that
\begin{eqnarray*}
\langle |\alpha_k(t)|^2 \rangle = \left( \sinh^2 \delta_{k_0} t + \cosh^2 \delta_{k_0} t \right) \langle |\alpha_k(0)|^2 \rangle.
\end{eqnarray*}

\end{document}